\documentclass[12pt]{article}

\usepackage{amssymb}
\usepackage{graphicx}

\begin{document}


\title{Properties of maximum Lempel-Ziv complexity strings}

\author{C. A. J. Nunes \\
\small Instituto de F\'isica, Universidade Federal de Uberl\^andia, 38408-100, Uberl\^andia, Brasil. \\
\and 
E. Estevez-Rams* \\
\small Facultad de F\'isica-IMRE, Universidad de La Habana, La Habana, Cuba. \\ \small estevez@imre.oc.uh.cu.
\and
B. Arag\'on Fern\'andez\\
\small Universidad de las Ciencias Inform\'aticas (UCI), La Habana, Cuba. \\
\and
R. Lora Serrano\\
\small Instituto de F\'isica, Universidade Federal de Uberl\^andia, 38408-100, Uberl\^andia, Brasil.}

\begin{abstract}
The properties of maximum Lempel-Ziv complexity strings are studied for the binary case. A comparison between MLZs and random strings is carried out. The length profile of both type of sequences show different distribution functions. The non-stationary character of the MLZs are discussed. The issue of sensitiveness to noise is also addressed. An empirical ansatz is found that fits well to the Lempel-Ziv complexity of the MLZs for all lengths up to $10^6$ symbols.
\end{abstract}



\date{\today}

\maketitle
\section{Introduction}

Lempel-Ziv complexity measure \cite{lz76} (from now on LZ76 complexity) has been used to analyze data sequences from different sources, theoretical or experimental \cite{szczepanski04,abo06,zhang09,contantinescu06,chelani11,rajkovic03,liu12,talebinejad11}. In an early paper, Ziv showed that the asymptotic value of the LZ76 complexity growth rate (LZ76 complexity normalized by $n/\log{n}$, where $n$ is the length of the sequence) is related to the entropy rate (as defined by Shannon information theory) for and ergodic source \cite{ziv78}. This theorem is essential at using the LZ76 complexity as an entropy rate estimator for finite sequences.

Entropy rate has a close relationship with algorithmic complexity, also known as Kolmogorov-Chaitin complexity. Algorithmic complexity measures the length of the shortest program, run in a Universal Turing Machine, that allows to reproduce the analyzed sequence. It was introduced as an attempt to avoid probabilistic arguments in analyzing single data sets \cite{calude02}. It is closely related to randomness, as infinite sequence with maximum algorithmic complexity are random . As random sequences have maximum entropy rate, it is often inferred that finite sequences with maximum LZ76 complexity are necessarily as random as they can be and viceversa \cite{li94}, yet neither of these two assumptions are true as was recently demonstrated \cite{estevez13}.

In an earlier paper \cite{estevez13}, the authors showed that from the algorithmic nature of the LZ76 factorization, there is a definite way to construct maximum LZ76 complexity sequences (we will call such sequences MLZs), that allows a much sorter description that the string itself. In this paper we will discuss the properties of such sequences.  

\section{Lempel-Ziv factorization and complexity}\label{sec:lz76}

Consider a sequence $u=u_1u_2\dots u_N$, where symbols are drawn from a finite alphabet $\Sigma$ of cardinality $\sigma(=|\Sigma|)$. Let $u(i,j)$ be the substring $u_iu_{i+1}\dots u_j$ taken from $u\;(u(i,j) \subset u)$ and of length j-i+1. It is understood that if $j>N$ we take up to the last character $u_N$ of $u$. If $i>j$ then $u(i,j)$ will be the empty string.

Let the ''drop'' operator $\pi$ be defined as

\begin{equation} 
u(i,j) \pi=u(i,j-1) \nonumber
\end{equation}

and, consequently,

\begin{equation}
u(i,j) \pi^k=u(i,j-k). \nonumber
\end{equation}

The Lempel-Ziv factorization\footnote{There are different Lempel-Ziv scheme for factorization \cite{cover06} and the reader must be careful to recognize which scheme is used in each case.} $E(u)$ of the string $u$ 

\begin{equation}
E(u)=u(1,h_1)u(h_1+1,h2)\dots u(h_{m-1}+1,N), \nonumber
\end{equation}

in $m$ factors is such, that each factor $u(h_{k-1}+1,h_k)$ complies with

\begin{enumerate}
 \item $u(h_{k-1}+1,h_k)\pi\subset u(1,h_k)\pi^2$
 \item $u(h_{k-1}+1,h_k)\not\subset u(1,h_k)\pi$ except, perhaps, for the last factor $u(h_{m-1}+1,N)$.
\end{enumerate}

The first condition defines $E(u)$ as a history of $u$, while the second condition defines such history as an exhaustive history of $u$. The partition $E(u)$ is unique for every string \cite{lz76}.

For example the exhaustive history of the sequence

$u=11011101000011$

is

$E(u)=1.10.111.010.00.011$

where each factor is delimited by a dot.

The LZ76 complexity $C(u)$ $(=|E(u)|)$ of the sequence $u$, is then defined as the number of factors in its exhaustive history. In the example above, $C(u)$=6.

In the limit of very large string length , $C(u)$ is bounded by \cite{lz76}

\begin{equation}\label{asymp}
 C(u) < \frac{N}{\log{N}}.
\end{equation}

It has been discussed \cite{estevez13} that this limit is attained for extremely long strings that can go beyond $10^{50}$ for the binary case, being even worse for larger alphabets (see below).

Let the entropy rate be given by 

\begin{equation}\label{shannonh}
 h= \lim_{N\rightarrow\infty}\frac{H[u(1,N)]}{N}
\end{equation}

where $H[u(1,N)]$ is the Shannon block entropy \cite{cover06} of the string $u(1,N)\in u$. Then, defining

\begin{equation}\label{lzh}
 c(u)=\frac{C(u)}{N/\log{N}}
\end{equation}

Ziv \cite{ziv78} proved that, if $u$ is the output from an ergodic source, then

\begin{equation}\label{zivtheorem}
 \limsup_{N\rightarrow\infty}c(u)=h.
\end{equation}

\section{Maximum Lempel-Ziv complexity sequences}\label{sec:mlzs}

For a given length $k$, there will be $\sigma^k$ different strings ($\sigma$ is the cardinality of the alphabet). Consider the following generating algorithm introduced in \cite{estevez13}

\begin{enumerate}
 \item Take $l=1$.

 \item Build the set $\Lambda^l$ of all strings of length $l$.\label{stepset}

 \item (Test step) Consider in lexicographic order an element  $\lambda \in \Lambda^l$, and take the string $u \lambda$ resulting from the concatenation of $u$ and $\lambda$. if $\lambda \not\subset u\lambda\pi$ append $\lambda$ to $u$, ignore otherwise.\label{teststep}

 \item Repeat step \ref{teststep} until all elements of $\Lambda^l$ have been considered or, $|u|\geq N$.

 \item If $|u| < N$, set $l=l+1$ and return to step \ref{stepset}.

 \item If $|u|>N$, truncate $u$ to length N from the end .

 \item Stop
\end{enumerate}

The algorithm will generate by construction a string of length $N$ of maximum Lempel Ziv complexity.

An example for an alphabet $\Sigma={a, b, c}$ will start by the alphabet symbols

\begin{equation}
 E(MLZs)=a.b.c \nonumber
\end{equation}

next, consider the set of all length two strings: ${aa, ab, ac, ba, bb, bc, ca, cb, cc}$. The first element of the
set will be a component of the exhaustive history of the sequence, so it will be append to the string

\begin{equation}
 E(MLZs)=a.b.c.aa \nonumber
\end{equation}

the second element in the set $ab$ is already present as the substring $u(1,2)$, thus, it is discarded. The third, fourth,
fifth and last element contributes to the exhaustive history

\begin{equation}
  E(MLZs)=a.b.c.aa.ac.ba.bb \nonumber
\end{equation}

Now turn to the length three set of strings and repeat the same procedure and so forth. It is clear that the above algorithm will yield a maximum LZ76 complexity string for a given string length $N$.

By construction, the maximum LZ76 complexity string is not unique. If we choose to test the candidate factor in any other order besides lexicographic order, a different sequence will happen. The set of all maximum LZ76 complexity sequences has been called MLZs.

\section{Properties of the MLZs}

\subsection{Bound values}

\begin{figure}
\includegraphics[scale=1.3]{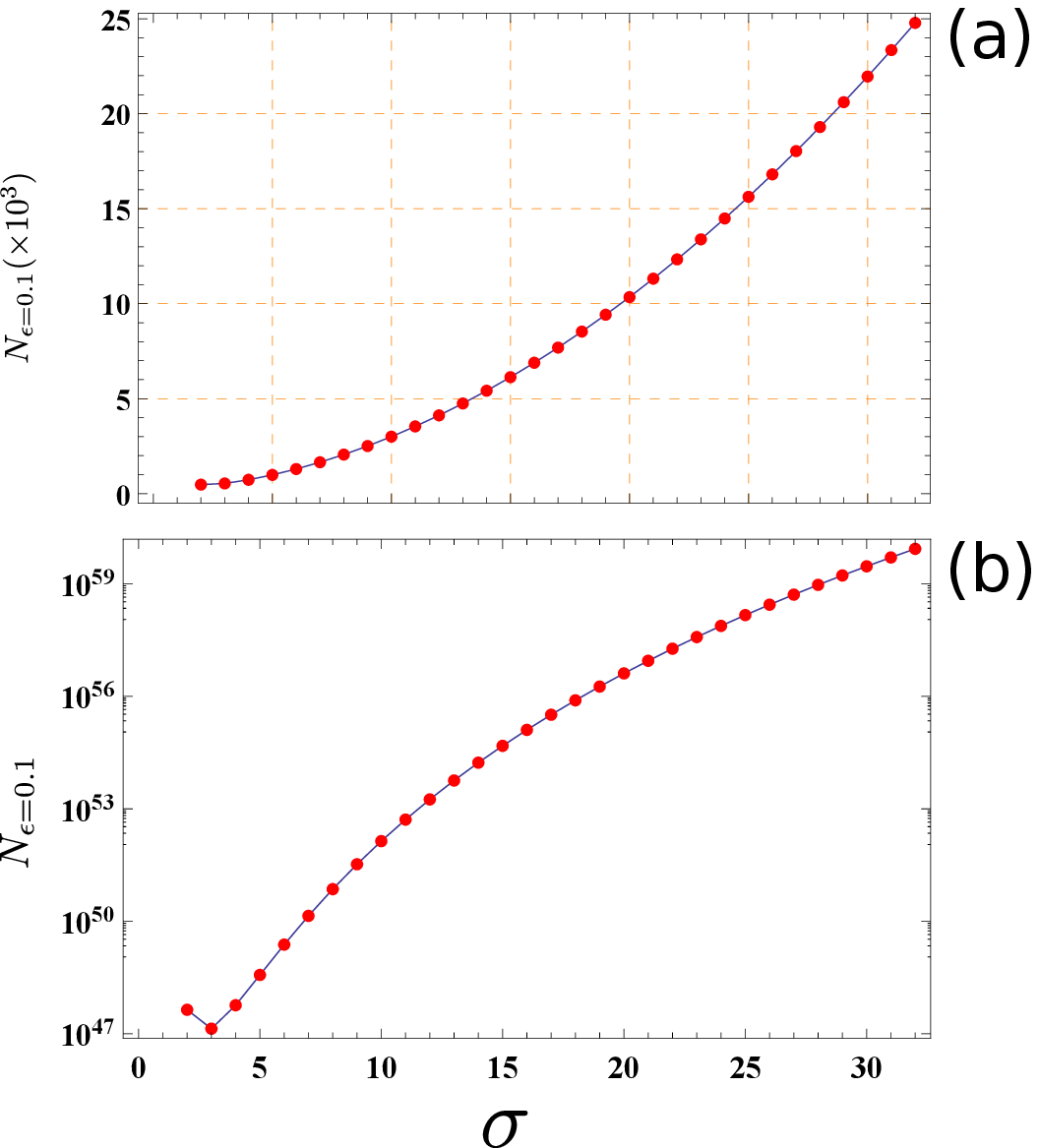}
\caption{(a) Sequence length $N_{\epsilon=1}$ for which $\varepsilon=1$, and (b) sequence length $N_{\epsilon=0.1}$ for which $\varepsilon=0.1$, as a function of alphabet size $\sigma$. Notice that (b) uses a log scale.}\label{epsilonfig}
\end{figure}

The asymptotic bounding value for the LZ76 complexity given by equation (\ref{asymp}), is deduced from the limit to the asymptotic function \cite{lz76}

\begin{equation}\label{epsilonasymp}
 C(u) < \frac{N}{[1-\epsilon(N)]\log{N}}.
\end{equation}

where

\begin{equation}\label{epsilon}
 \epsilon(N) = 2\frac{1+\log_{\sigma}{\log_{\sigma}{\sigma N}}}{\log_{\sigma}{N}}.
\end{equation}

is a slowly decaying function of $N$.  It should be emphasized that one problem with equation (\ref{epsilonasymp}) is that its interval of validity is given by the condition $\epsilon(N)<1$, which is not the case for every value of $N$. Figure \ref{epsilonfig}a shows the value of $N$ for which $\epsilon(N)=1$ for increasing alphabet size $\sigma$. For an alphabet size of $26$, the validity is already in the order of $10^4$. Even worse, $\epsilon$ is  a very slowly decaying function of $N$, figure \ref{epsilonfig}b plots the value of the $N$ for which  $\epsilon(N)=0.1$ for increasing alphabet size. Already for the binary alphabet, the value of $N$ is of the order of $10^{50}$, and can reach $10^{60}$ for alphabet size $25$ (cardinality of western languages alphabets). 

\begin{figure}
\includegraphics{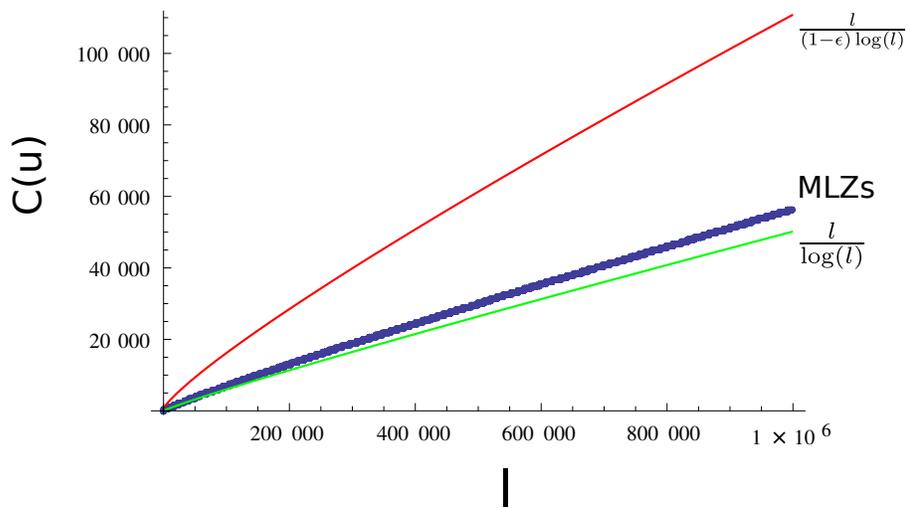}
\caption{Maximum Lempel-Ziv string LZ76 complexity as a function of string length. The MLZs curve lies between the asymptotic values given by equations (\ref{asymp}) and (\ref{epsilonasymp}).}\label{epsilonmlzs}
\end{figure}

The bound value given by equation (\ref{asymp}) is only reached for extremely large strings and on the other hand, the bound value given by equation (\ref{epsilonasymp}) is, for small, and not so small, values of $N$ a to gross approximation. Indeed, Figure \ref{epsilonmlzs} shows the LZ76 complexity of the MLZs, equation (\ref{asymp}) and equation (\ref{epsilonasymp}). It is evident that while the asymptotic value is not an upper bound for the LZ76 complexity up to $10^6$ symbols, the bound given by equation (\ref{epsilonasymp}) is well above the maximum LZ76 complexity attainable.

\subsection{Non-randomness of MLZs}

MLZs are non-random strings by construction, this is probably its most striking property. Ziv theorem \cite{ziv78}, given by equation (\ref{zivtheorem}), could wrongly induce to infer that random strings of finite length should achieve the maximum possible LZ76 complexity for that length. This is not the case as has been already reported \cite{estevez13}. 

A truly random source will output any string of length $N$ with probability $\sigma^{-N}$, without regard for the actual string being produced. A typical random string, will be given by any string that can pass the universal Martin-L\"of test of randomness \cite{martin66}. Such string will have several properties one of which, is maximum algorithmic complexity as well as, normality in Borel sense \cite{li94}. A sequence is considered normal in Borel sense, if the probability of occurrence of all $k$-length patterns are equal, and tend to $\sigma^{-k}$ for infinite strings.  It has been reported that MLZs are not normal in Borel sense \cite{estevez13}.

\begin{figure}
\includegraphics{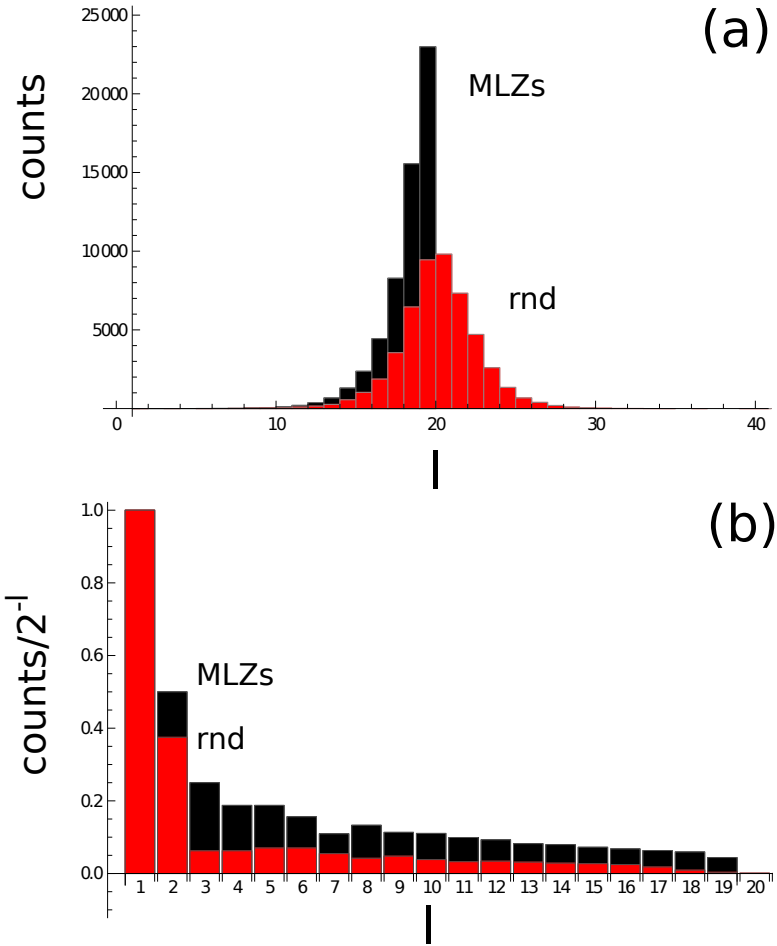}
\caption{Length distribution for the LZ76 factors for the MLZs and the random string. (a) The histogram of the length distribution of the factors in the exhaustive history of the MLZs and the random (rnd) strings. (b) Normalized distribution. Each count value was normalized by $2^{-l}$ where $l$ is the factor length. For both plots, each sequence had a total size of $10^5$ symbols, the calculation was performed 100 times and average values were taken.}\label{histolength}
\end{figure}

Louchard and Szpankowski \cite{louchard94} has studied the probability distribution of factors length in a Lempel-Ziv factorization slightly different from the one studied in this contribution\footnote{Louchjard and Szpankowski use the factorization defined by Lempel and Ziv in a paper in 1978 \cite{lz78} which is a modified scheme of LZ76.}.They found that for a biased Bernoulli model (symbols are drawn from a binary alphabet with different probability $\alpha$ and $1-\alpha$), the distribution of length follows a Gaussian law. Figure \ref{histolength}a shows the length distribution for both the random sequence factorization, and the MLZs factorization using the LZ76 scheme described above. The string length was taken as $10^5$ and $1000$ strings for each case were taken. Results are from the average values. The random string shows a Gaussian distribution with mean length value of $19.7$ and standard deviation of $2.5$. The MLZs, on the other hand, exhibit a length distribution far from being Gaussian with an sudden cut-off length value of $19$, where the number of factors of length above this value falls abruptly. We normalize the number of factors for each length $l$, by the total number of strings of that given length $2^{l}$, the results is shown in  figure \ref{histolength}. Each value corresponds to the fraction of factors of length $l$ that actually comes into the factorization. The behavior for both the random and the MLZs strings are similar in nature, decreasing with factor length. For all lengths, the fraction of factors in the random strings are smaller that those for the MLZs. 
W
e can calculate a Shannon entropy related magnitude over the length distribution as

\begin{equation}
 H(u)=-\sum_{l=1}^{\infty}p_{l}\log_{2}p_{l}
\end{equation}

where $p_{l}$ is a probability defined by the fraction of factors of length $l$ in the LZ factorization of $u$, relative to the total number of different binary strings of length $l$ ($2^{l}$). It must be noted that this is not strictly a true Shannon entropy, as the sum of the probabilities $p_{l}$ are not normalized.

For the random string this gives a value $H(rnd)=3.31$ compared with  $H(MLZs)=2.25$ for the MLZs. There is more surprise, on average, in the length distribution of the random string that the one found in the MLZs, which corresponds to less uniform distribution of the MLZs compared to the random case.

\subsection{Sensibility to the order of the testing step.}

\begin{figure}
\includegraphics{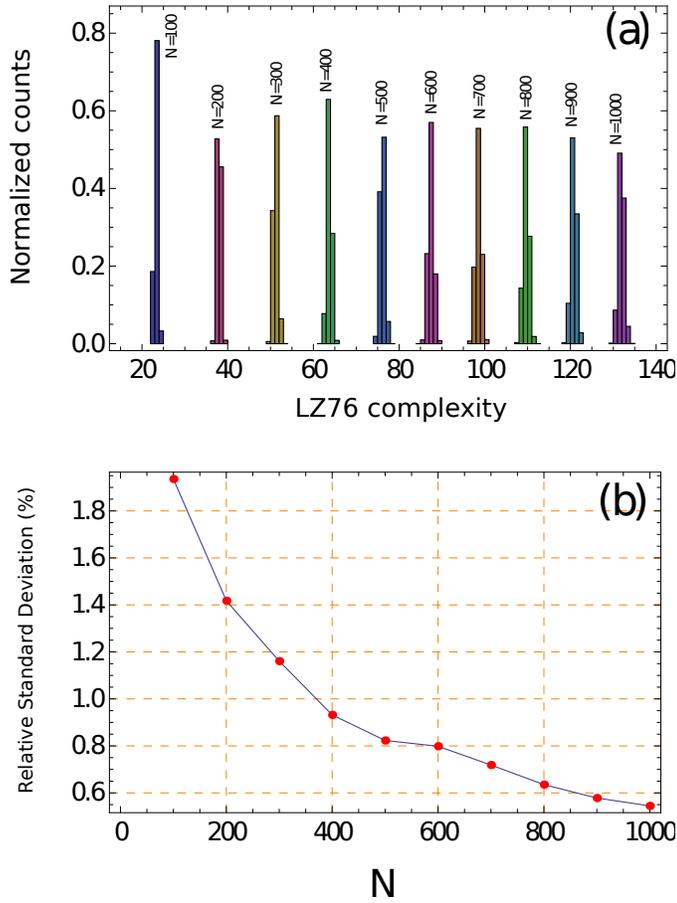}
\caption{Dispersion of LZ76 complexity values for the MLZs. (a)Histogram of LZ76 complexity values for different sequence length $N$. The MLZs where built $5\times 10^3$ times for each length using random ordering in the testing stage. (b) Relative standard deviation of the LZ76 complexity values as a function of $N$.}\label{histomlzs}
\end{figure}

It may be asked if the complexity of the resulting MLZs are significantly sensitive to the ordering chosen in the
testing step, this seems not to be the case. Five thousand sequences were built using random ordering in the test step.
Figure \ref{histomlzs}a shows the histogram of the LZ76 complexity value for length ranging from $10^2$ to $10^3$
symbols, it can be seen that the dispersion of values is very narrow for each length. Figure \ref{histomlzs}b shows that
the relative standard deviation falls rapidly with increasing sequence length, being already below $2\%$ for the $10^2$
symbol sequences.

\subsection{Non-stationary character}

\begin{figure}
\includegraphics{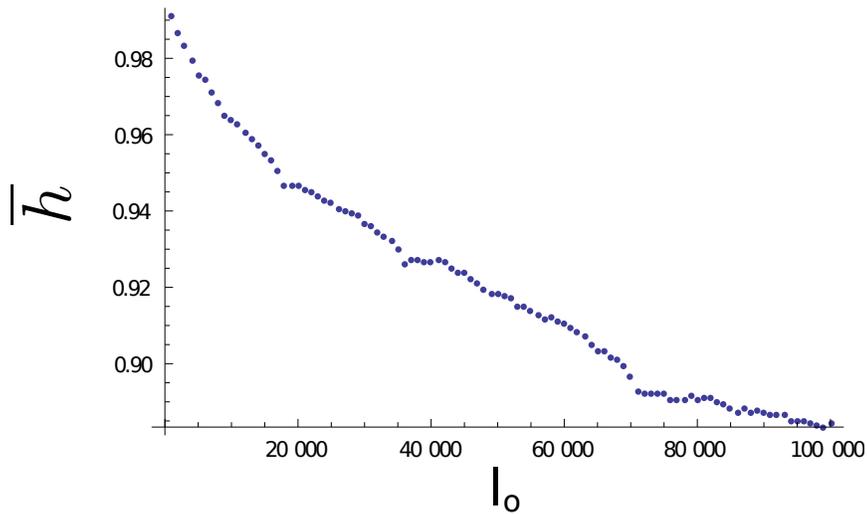}
\caption{Entropy rate $\overline{h}$ from the normalized LZ76 complexity for the MLZs strings as a function of the starting index $l_{o}$, where the factorization begins. Entropy rate was estimated by normalizing the LZ76 complexity for the full length sequence, by the same length complexity of the original MLZs. The MLZs strings had it first $l_{o}$ symbols dropped before LZ76 factorization was performed. The variation in the LZ76 complexity points to the non stationary character of the strings. Each sequence had a total size of $10^5$ symbols, the calculation was performed $100$ times and average values were taken.}\label{stationarity}
\end{figure}

The MLZs are by construction non stationary and non random. The non stationary character of the MLZs is result of the construction procedure. How sensitive are the MLZs to the starting point for factorization can be seen in figure \ref{stationarity}. The figure plots the estimated entropy rate for the MLZs, starting the factorization at different positions of the MLZ string. All strings where $10^5$ length. The entropy rate was estimated from the LZ76 complexity value for the whole string, normalized by the LZ76 complexity of the original MLZs. It can be seen that the entropy rate drops as we factorize starting further along the MLZs strings.

\subsection{Sensibility to noise}

We corrupted the MLZs strings with white noise in order to study their sensibility to noise. A fraction $\alpha$ of sites chosen randomly were flipped in the original MLZs. For such corrupted vector of values, the LZ76 factorization was carried out and the LZ76 complexity calculated. The entropy rate was estimated using the LZ76 normalization. For each $\alpha$ value, $100$ strings were used for the calculation and averaged values are reported.
 
\begin{figure}
\includegraphics{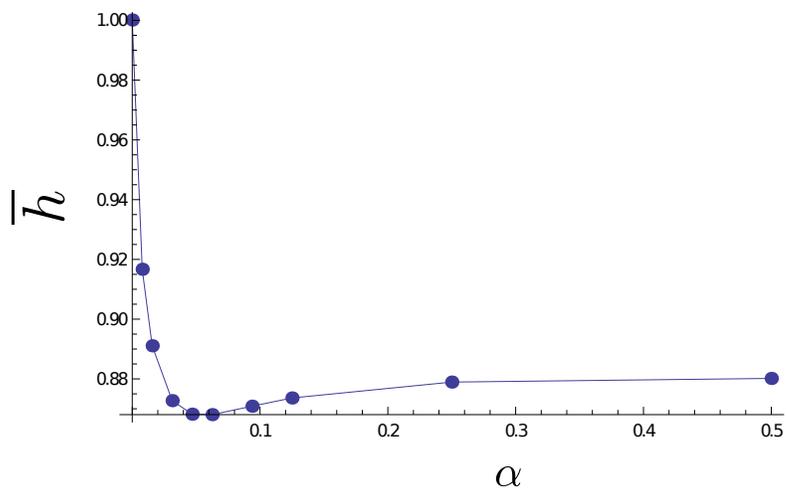}
\caption{Entropy rate $\overline{h}$ for the MLZs strings as a function of percentage of white noise. Entropy rate was estimated as in figure \ref{stationarity}. $\alpha$ is the fraction of sites in the string where a flip of the value was performed. The flipped sites were taken randomly.  Each sequence had a total size of $10^5$ symbols, the calculation was performed $100$ times and average values were taken.}\label{mlzsnoise}
\end{figure}

Figure \ref{mlzsnoise} shows the behavior of the estimated entropy with percentage of noise. There is an abrupt fall of entropy rate up to $\alpha=0.06$ and then, a slight increase followed by a certain stabilization can be observed.

\subsection{Functional behavior of MLZs}

\begin{figure}
\includegraphics{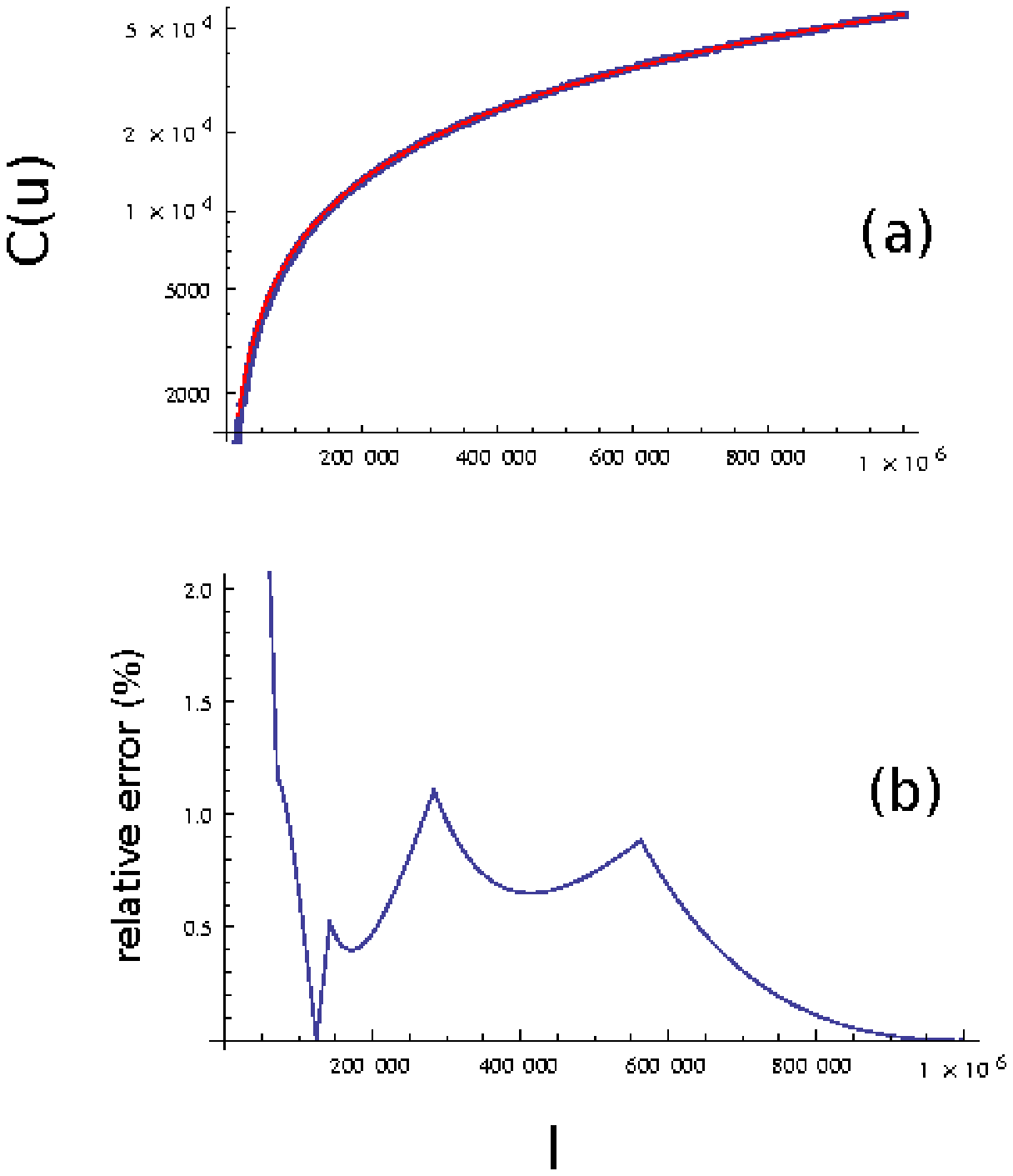}
\caption{(a) The LZ76 complexity C(u) (in blue) and the fitted empirical ansatz given by equation (\ref{eq:ansatz} (in red). (b) The absolute  relative error between the actual data and the fitted values.}\label{fitmlzs}
\end{figure}

Finally, the MLZs was modeled using the empirical ansatz

\begin{equation}
 C(u_{n})= b \frac{n^{c}}{(1-\epsilon(n)) \log(n)}\label{eq:ansatz}
\end{equation}

where  $b$, $c$ are fitting parameters. The $R^2$ value for the fit was nearly one ($0.9999$). The fitted $c$ value was $1.06$, very close to $1$ which corresponds to the expected asymptotic value, while $b$ was fitted to $0.153$. Figure \ref{fitmlzs}a shows the agreement of the fitted function with the actual data in a log scale. The absolute relative errors of the fitted function with the data values (residuals) are shown in figure \ref{fitmlzs}b, the relative errors do not exceed the $1.1\%$ for string lengths above $150000$ symbols.

\section{Summary}
Maximum Lempel-Ziv complexity sequence are far from random and can be generated by a deterministic algorithm of size $O(\log n)$ which makes them of negligible algorithmic complexity, exactly the opposite to a random string. MLZs are mostly insensitive to the order in which the possible factors in the exhaustive history are tested. MLZs do not show normality in Borel sense and the factor length distributions departs markedly from the Gaussian distribution with an abrupt cut-off value for large enough factors. Compared to the random sequence, the MLZs have smaller Shannon entropy of the length distribution. The analytical asymptotic values in used, are only valid for very large sequences that can go up to $10^{50}$ symbols. MLZs are strongly non stationary and a gradually loss of entropy rate happens as  we start factorization dropping larger initial segments of the sequence. The same behavior was observed when the MLZs were corrupted with white noise, in this case an abrupt drop of entropy rate down to 0.89 happened, when only 0.06 fraction of sites were flipped.

We finally introduced an empirical ansatz for describing the LZ76 complexity of the MLZs which behaved very good in the whole range of sequence lengths, with relative error not surpassing 2\%.

\section{Acknowledgments}

E. Estevez-Rams and RLS which to thanks FAPEMIG (BPV-00039-12) for financial assistance in mobility. Universidad de la Habana and Universidad de la Ciencias Inform\'aticas are acknowledge for partial financial support and computational infrastructure.


\pagebreak

\end{document}